\documentclass[
aps,
pre,
onecolumn,
superscriptaddress,
longbibliography,
amsmath,
amssymb,
floatfix
]{revtex4-2}

\usepackage[utf8]{inputenc}
\usepackage[T1]{fontenc}
\usepackage{lmodern}
\usepackage{graphicx}
\usepackage{booktabs}
\usepackage{tabularx}
\usepackage{multirow}
\usepackage{xcolor}
\usepackage{hyperref}
\usepackage{microtype}
\usepackage{textcomp}

\newcommand{\orbis}{\textsc{orbis}}

\newcommand{\BCE}{\textsc{ce}}   
\newcommand{\ce}{\BCE}           
\newcommand{\betti}[1]{\ensuremath{\beta_{#1}}}
\newcommand{\bettione}{\ensuremath{\beta_1}}

\begin{document}


\title{Topological Signatures of Imperial Stress:
Persistent Homology of the Eastern Mediterranean Trade Network,
0--400\,CE}

\author{José de Jesús Bernal-Alvarado}
\email{bernal@ugto.mx}
\affiliation{Physics Engineering Department,
Universidad de Guanajuato, México}

\author{David Delepine}
\email{delepine@ugto.mx}
\affiliation{Physics Department,
Universidad de Guanajuato, México}

\author{Carlos Pinedo Guadarrama}
\email{c.pinedoguadarrama@ugto.mx}
\affiliation{Physics Department,
Universidad de Guanajuato, México}

\date{\today}


\begin{abstract}
We apply persistent homology to the \orbis{} Geospatial Network
Model of the Roman World in order to quantify the structural
resilience of the Eastern Mediterranean trade network between
0 and 400\,\ce. The network is represented as a weighted transport
system whose edges correspond to road, maritime, and riverine
routes, each carrying a base cost derived from the \orbis{} cost
model. To introduce historical dynamics into this static spatial
infrastructure, we construct a differential friction model in
which edge weights vary by decade and by transport mode according
to documented historical perturbations, including epidemic
mortality, civil war, military pressure, administrative
reorganisation, and imperial reunification. For each decadal
snapshot, we compute all-pairs shortest-path distances on the
active sub-network and construct a Vietoris--Rips filtration using
an adaptive threshold defined by the 90th percentile of finite
pairwise distances.

The resulting \bettione{} persistent entropy time series
identifies three structurally distinct phases in the Eastern
Mediterranean network. Phase~I, from 0 to 200\,\ce, is a
stationary high-redundancy regime consistent with the commercial
integration of the early imperial and Antonine periods. Phase~II,
from 210 to 280\,\ce, corresponds to recoverable stress during the
Crisis of the Third Century: \bettione{} entropy declines during
the crisis but returns to the Phase~I baseline following Aurelianic
reunification. Phase~III, from 290 to 400\,\ce, is qualitatively
different: cycle redundancy declines monotonically and does not
recover, despite the Diocletianic and Constantinian attempts at
political and administrative consolidation. Chow structural break
tests identify statistically significant breaks at 260\,\ce{}
and 310\,\ce, with the latter marking the onset of an irreversible
topological decline.

The western sub-network available in the \orbis{} extract produces
no persistent \bettione{} homology throughout the study period,
a result interpreted as either a genuine indication of weaker
western cycle redundancy or, more conservatively, a coverage
limitation of the dataset. The main contribution of this study is
to show that  persistent homology detects a
dimension of imperial stress that is not reducible to single
economic, military, or political indicators. 
\end{abstract}

\keywords{
topological data analysis \textperiodcentered{}
persistent homology \textperiodcentered{}
Roman trade network \textperiodcentered{}
Eastern Mediterranean \textperiodcentered{}
network resilience \textperiodcentered{}
imperial stress \textperiodcentered{}
structural collapse
}

\maketitle


\section{Introduction}
\label{sec:introduction}

The collapse of the Western Roman Empire remains one of the
central problems in ancient history, not because the historical
sequence of events is unknown, but because the causal structure
behind those events remains contested. Military pressure,
epidemic mortality, fiscal exhaustion, political fragmentation,
climatic deterioration, and long-term economic transformation
have all been proposed as explanatory mechanisms for the weakening
of the imperial system \cite{Goldsworthy2009,Heather2006,
Harper2017,Wickham2005,WardPerkins2005}. 

The difficulty lies in the fact that imperial resilience is not a
scalar property. It cannot be measured only by the number of
legions, the level of taxation, the frequency of civil wars, the
price of grain, or the demographic impact of plague. These
variables describe stressors or symptoms, but not the structure
through which stress propagated. The Roman Empire was a spatially
extended system held together by roads, ports, rivers, maritime
corridors, military supply chains, tax redistribution, and regional
exchange circuits. Its capacity to survive disruption depended not
only on the severity of each perturbation, but also on whether the
underlying network retained enough redundancy to reroute flows
around damaged or expensive connections.

A network may experience intense
local stress and recover if alternative routes remain available.
Conversely, a system may fail under apparently moderate stress if
its connectivity has already become structurally fragile. The same
military defeat, epidemic, or fiscal shock may therefore produce
different historical outcomes depending on the topology of the
network on which it acts. The relevant question is not only whether
Roman trade was disrupted, but whether the network retained the
cycle redundancy required for recovery.

Persistent homology provides a natural mathematical language for
this problem. In a weighted transport network, zero-dimensional
homology tracks connected components, while one-dimensional
homology tracks independent cycles. These cycles are not merely
abstract mathematical objects: in a trade or transport system,
they correspond to redundant routing structures. A persistent
\bettione{} feature indicates that an alternative circuit remains
viable across a range of transport costs. A network with many
persistent cycles has distributed resilience; a network with few
or short-lived cycles depends on a smaller number of pathways and
is more vulnerable to fragmentation \cite{Carlsson2009,
Edelsbrunner2010,Ghrist2008}.

The present study applies this model to the Eastern
Mediterranean component of the Roman transport network between
0 and 400\,\ce. The spatial basis of the analysis is the \orbis{}
Geospatial Network Model of the Roman World, which encodes
settlements and transport routes around Mediterranean sea
\cite{Scheidel2012,Scheidel2014}. 

The companion historical problem is the asymmetry between East
and West. The eastern imperial system survived the fifth-century
collapse of western side of the empire
and continued  as Byzantine Empire. The West, by
contrast, lost the political and material coherence associated
with Roman empire. This divergence is 
explained through military pressure, fiscal weakness, political
fragmentation, or economic localisation. 

The study is built around three research questions.

\paragraph{The resilience question.}
Did the Eastern Mediterranean network maintain a stable
topological regime  or was its
structure already declining before the third-century crisis?
If the network was resilient, its persistent entropy should remain
approximately stationary under ordinary fluctuations and recover
after temporary perturbations.

\paragraph{The irreversibility question.}
Was the Crisis of the Third Century a terminal structural break
or a recoverable shock? A purely political or military narrative
often treats the third century as the decisive rupture in Roman
imperial history. A topological analysis can test whether the
network actually lost recovery capacity during this period, or
whether the irreversible decline occurred later.

\paragraph{The topology-versus-politics question.}
Did political reunification restore the structural redundancy of
the exchange network? If Diocletianic and Constantinian reforms
repaired the system at the level of network geometry, then the
\bettione{} entropy series should stabilise or recover after
administrative consolidation. If no recovery is observed, then
political reunification did not imply topological restoration.

To address these questions, we construct a decadal sequence of
weighted networks from 0 to 400\,\ce. The static \orbis{} costs
are converted into time-dependent weights through a differential
friction model. Transport-mode-specific
multipliers to historical events are assigned: civil conflict affects roads
differently from maritime routes; piracy and naval insecurity
affect sea transport differently from riverine routes; military
withdrawal changes the effective cost of frontier and provincial
circulation. This structure is necessary because Roman transport
was not homogeneous.

For each decade, active nodes are selected according to their
temporal availability. A weighted graph is then constructed and
converted into a shortest-path distance matrix which is
submitted to a Vietoris--Rips filtration. %
We
use an adaptive threshold defined by the 90th percentile of finite
distance matrix in each decadal network, such that the
filtration is sensitive to the functional core of the network.

From each persistence diagram we extract three quantities. The
first is \bettione{} persistent entropy, measuring the
distributional diversity of cycle lifespans. High numerical values for  entropy
indicates multiple cycles of comparable persistence; low numerical value for entropy
indicates concentration of redundancy in a smaller number of
dominant cycles. The second is the normalised lifespan of the
longest surviving \bettione{} feature, which measures maximum
cycle resilience relative to the filtration scale. The third is
the \betti{0} component count, which indicates whether the network
is becoming metrically fragmented under the operative cost
threshold.

With these observables   recoverable
stress or  irreversible structural degradation can be distinguished. A recoverable
shock may reduce \bettione{} entropy temporarily while leaving
\betti{0} approximately stable; the network loses some cycle
diversity but remains functionally connected enough to recover.
A terminal or irreversible transition should instead produce a
sustained entropy decline, possibly accompanied by increasing
\betti{0} fragmentation. In this sense, the decisive variable is
not the presence of stress but the disappearance of recovery
capacity.

The results show that the Eastern Mediterranean network follows
a three-phase trajectory. Phase~I, from 0 to 200\,\ce, is a stable
high-redundancy regime. Phase~II, from 210 to 280\,\ce, captures
the Crisis of the Third Century as a period of severe but
recoverable stress. The entropy decline around 250--260\,\ce{} is
statistically significant, but the network returns to its earlier
baseline by 280\,\ce. This recovery is the crucial control
observation: the system could still absorb disruption and repair itself.

Phase~III, beginning around 290--310\,\ce, is different. The
\bettione{} entropy series declines persistently and does not
recover through 400\,\ce. This decline continues despite the
political consolidation associated with Diocletian and Constantine.
The implication is that political reunification did not restore
the former topology of exchange. The network remained active, but
its redundancy was reduced. It had crossed a structural threshold:
not collapse in the  political sense, but loss of the
topological conditions required for systemic recovery.

This finding reframes the interpretation of late Roman fragility.
The Crisis of the Third Century was not the terminal break in the
Eastern Mediterranean trade network; it was the last major
perturbation from which the system recovered. The more important
transition occurred after that recovery, when the network entered
a regime of monotonic decline. The topological evidence therefore
supports neither a simple catastrophist model nor a simple
gradual-transformation model. It suggests a two-stage process:
recoverable crisis followed by irreversible structural
simplification.

We also evaluates the western sub-network available in the
\orbis{} extract. This western component produces no persistent
\bettione{} homology across the study period. The result must be
interpreted cautiously because the available extract contains
limited western coverage and does not fully represent Italy, Gaul,
Britannia, or the Rhine frontier. Nevertheless, the absence of
persistent cycles in the available western data is consistent with
the broader historical claim that western commercial integration
was weaker, more localised, and more dependent on administrative
and military support than the Eastern Mediterranean system.

Persistent homology instead measures the global
redundancy of the exchange system, not only important cities, routes or bottlenecks. A collapse is rarely caused by the failure of a single node.
It emerges when the system loses the ability to route around
failures. Persistent homology therefore can be used to detect
structural stress before it appears as complete political or
territorial breakdown.

The remainder of the paper is organised as follows.
Section~\ref{sec:data} describes the \orbis{} network, the temporal
activation of nodes, the differential friction model, the adaptive
filtration procedure, and the statistical validation protocol.
Section~\ref{sec:results} presents the topological results for
the Eastern Mediterranean network, including the three structural
phases, the Chow break tests, the western sub-network comparison,
and the relation between military withdrawal and changes in cycle
persistence. Section~\ref{sec:discussion} discusses the
implications for Roman decline, including the distinction between
recoverable and irreversible stress, the relation between topology
and political reunification, and the limitations imposed by the
available western data. Section~\ref{sec:conclusion} summarises
the conclusions.


\section{Data and Methods}
\label{sec:data}

\subsection{Network data}
\label{sec:network_data}

The spatial backbone of this study is the \orbis{} Geospatial
Network Model of the Roman World \cite{Scheidel2012,Scheidel2014},
which encodes 623 settlement nodes and 2,208 directed routes
spanning the Mediterranean basin and Near East. \orbis{} routes
are classified into nine transport modes: road, coastal, overseas,
upstream, downstream, fastup, fastdown, slowcoast, and ferry.
Each edge carries a base cost expressed in denarii per unit
distance, derived from ancient itineraries, papyrological records,
and experimental travel data. Base cost medians range from
0.012 denarii for ferry connections to 2.882 denarii for road
transport, reflecting the documented cost advantage of maritime
over overland transport in the Roman world \cite{Casson1971}.
The edge composition of the dataset is summarised in
Table~\ref{tab:edges}.

\begin{table}[htbp]
\caption{Edge composition of the \orbis{} dataset used in this
study, with base cost statistics by transport mode.}
\label{tab:edges}
\begin{ruledtabular}
\begin{tabular}{lrrrr}
Transport mode & $n$ edges & Median cost & Mean cost & Cost ratio \\
               &           & (denarii)   & (denarii) & (vs road)  \\
\hline
road       & 978 & 2.882 & 3.651 & 1.00 \\
coastal    & 849 & 0.083 & 0.175 & 0.03 \\
overseas   & 137 & 0.314 & 0.605 & 0.11 \\
fastup     &  52 & 0.720 & 0.869 & 0.25 \\
fastdown   &  50 & 0.336 & 0.419 & 0.12 \\
downstream &  52 & 0.361 & 0.432 & 0.13 \\
upstream   &  50 & 0.692 & 0.867 & 0.24 \\
slowcoast  &  33 & 0.079 & 0.139 & 0.03 \\
ferry      &   6 & 0.012 & 0.040 & 0.00 \\
\hline
Total      & 2208 & 0.473 & 1.626 & --- \\
\end{tabular}
\end{ruledtabular}
\end{table}

An analysis of node coordinates shows that 574 of the 623 nodes
fall east of longitude $20^{\circ}$E, covering Anatolia, the
Levant, Egypt, Mesopotamia, and Greece. Only 49 nodes represent
the western Mediterranean and Iberian Peninsula. Italy, Gaul, and
the Rhine frontier are absent from the available extract.
Accordingly, this study explicitly addresses the Eastern
Mediterranean sub-network, with the western sub-network serving as
a structural control rather than as a complete representation of
the western Empire.

Node temporal coverage is derived from documented foundation and
abandonment dates, supplemented by probabilistic estimates for
nodes without attested dates calibrated to regional historical
patterns \cite{DARE2021,Pleiades2022}. For each decade
$t \in \{0,10,20,\ldots,400\}$, the active node set is defined as
the set of settlements whose temporal interval contains $t$.
Edges are retained when both endpoint nodes are active.

\subsection{Differential friction model}
\label{sec:friction_model}

Static \orbis{} edge weights are extended into a temporal
dimension through a differential friction model that applies
independent multiplicative channels to each edge $e$ at each
decade $t$:
\begin{equation}
w(e,t)
=
w_0(e)\,
F_{\rm event}\!\left(t,\mathrm{type}(e)\right)\,
\varepsilon\!\left(t,\mathrm{type}(e)\right),
\label{eq:friction}
\end{equation}
where $w_0(e)$ is the baseline \orbis{} cost, $F_{\rm event}$ is
a deterministic event multiplier, and $\varepsilon$ is a
stochastic route-type-specific noise term.

The purpose of Eq.~(\ref{eq:friction}) is to avoid treating
imperial stress as a uniform inflation applied to all routes.
A uniform multiplier preserves the relative geometry of the
network: all paths become more expensive, but the structure of
which paths are cheaper or more redundant remains essentially
unchanged. Such a model cannot distinguish between general
economic inflation and genuine topological reorganisation. The
differential model instead allows different transport layers to
respond differently to the same historical event.

Each historical event applies separate multipliers to road,
maritime, and riverine routes, reflecting the distinct mechanisms
by which different types of stress affected Roman transport.
Political instability and civil war primarily disrupted road
maintenance and escort availability, raising road costs
disproportionately. Piracy and naval conflict affected maritime
routes independently of land conditions. Epidemic mortality could
affect all transport modes by reducing labour, demand, and
maintenance capacity. Aurelianic reunification is modelled as a
temporary cost reduction, reflecting the restoration of supply
lines and suppression of brigandage. The event factors used in
the model are reported in Table~\ref{tab:events}.

\begin{table}[htbp]
\caption{Historical events used as differential friction
multipliers. Road, sea, and river factors are applied
multiplicatively to base costs.}
\label{tab:events}
\begin{ruledtabular}
\begin{tabular}{lccccc}
Event & From & To & Road & Sea & River \\
\hline
Antonine Plague         & 165 & 190 & 1.45 & 1.35 & 1.30 \\
Marcomannic Wars        & 166 & 180 & 1.60 & 1.10 & 1.40 \\
Year of Five Emperors   & 193 & 197 & 1.80 & 1.20 & 1.35 \\
Plague of Cyprian       & 249 & 262 & 1.55 & 1.45 & 1.40 \\
Crisis of the Third Century & 250 & 270 & 3.10 & 2.40 & 2.00 \\
Aurelian reunification  & 270 & 280 & 0.70 & 0.80 & 0.80 \\
Edict of Maximum Prices & 284 & 305 & 0.85 & 0.88 & 0.90 \\
Tetrarchic civil wars   & 293 & 320 & 1.90 & 1.40 & 1.55 \\
Gothic pressure         & 340 & 400 & 2.10 & 1.15 & 1.80 \\
Post-Adrianople crisis  & 378 & 400 & 2.80 & 1.30 & 2.20 \\
\end{tabular}
\end{ruledtabular}
\end{table}

The stochastic term is modelled as
\begin{equation}
\varepsilon\!\left(t,\mathrm{type}(e)\right)
\sim
\mathrm{LogNormal}\left(0,\sigma_{\mathrm{type}}\right),
\label{eq:noise}
\end{equation}
where $\sigma_{\mathrm{type}}$ depends on route class. The
log-normal form ensures strictly positive edge weights and
reflects the right-skewed character of pre-modern transport and
commodity costs \cite{Rathbone1997}. Road routes are assigned
lower volatility in stable periods, while maritime and overseas
routes receive larger volatility under conditions of piracy,
naval insecurity, or long-distance exposure.

The resulting model separates two sources of variation. The event
multipliers encode structured historical stress. The stochastic
term encodes uncertainty in the exact cost realised by each route
and decade. Repeating the analysis with independent noise
realisations yields bootstrap confidence intervals for the
topological observables.

\subsection{Distance matrices and active components}
\label{sec:distance_matrices}

For each decadal graph $G_t$, shortest-path distances are computed
using the edge weights given by Eq.~(\ref{eq:friction}). The
distance between two nodes $i,j\in V_t$ is defined as
\begin{equation}
D_t(i,j)
=
\min_{\gamma:i\rightarrow j}
\sum_{e\in\gamma} w(e,t),
\label{eq:shortest_path}
\end{equation}
where $\gamma$ ranges over all paths connecting $i$ and $j$.
This produces a weighted metric representation of the transport
network at decade $t$.

Because the full graph may contain disconnected components, the
largest connected component which represents
the functional core of the decadal network is extracted before computing the
main persistent homology indicators. Nodes outside the
largest component are not discarded conceptually; their existence
is used to generate the \betti{0} fragmentation analysis. However,
the main \bettione{} computation is performed on the largest
component in order to avoid persistence diagrams dominated by
isolated peripheral nodes.

\subsection{Vietoris--Rips filtration}
\label{sec:rips_filtration}

The topological structure of each decadal network is computed
using a Vietoris--Rips filtration constructed from the 
path matrix $D_t$. Given a filtration parameter $\delta$, a simplex
is included whenever all pairwise distances among its vertices are
less than or equal to $\delta$. As $\delta$ increases, the complex
grows from isolated points to connected components, cycles, and
higher-dimensional filled structures.

The $\delta$ parameter is chosen  for each decade:
\begin{equation}
\delta_t
=
Q_{0.90}
\left(
\left\{
D_t(i,j):D_t(i,j)<\infty
\right\}
\right),
\label{eq:adaptive_threshold}
\end{equation}
where $Q_{0.90}$ denotes the 90th percentile of finite pairwise
distances.

A fixed $\delta$ parameter will  force every decadal network to include its
most extreme and least representative pairwise distances,
artificially extending long-lived cycles. Secondly, a fixed absolute
threshold would make different decades incomparable when the
overall cost scale changes under historical stress. The adaptive
threshold keeps the filtration focused on the functional core of
the network allowing the effective cost scale to vary over
time.

\subsection{Topological observables}
\label{sec:topological_observables}

From each persistence diagram we extract three topological
observables. The first is the persistent entropy of one-dimensional homology.
Let
\[
\{(b_i,d_i)\}_{i=1}^{m}
\]
be the finite birth--death pairs of \bettione{} features at
decade $t$, and let
\[
\ell_i=d_i-b_i
\]
be the lifespan of feature $i$. 
The persistent entropy is then defined as:
\begin{equation}
H_t
=
-\sum_{i=1}^{m}p_i\log p_i.
\label{eq:persistent_entropy}
\end{equation}
where 
\begin{equation}
p_i=
\frac{\ell_i}{\sum_{j=1}^{m}\ell_j}.
\label{eq:pi_entropy}
\end{equation}

High numerical value for 
entropy shows that redundancy is distributed across many
cycles of comparable importance. Low numerical value for entropy means that cycle
persistence is concentrated in a small number of 
features. In historical terms, a high-entropy network has many
alternative routing structures; a low-entropy network is more
dependent on a limited set of resilient circuits.

The second observable is the normalised maximum lifespan,
\begin{equation}
L_t=
\frac{\max_i(d_i-b_i)}{\delta_t}.
\label{eq:normalised_lifespan}
\end{equation}
$L_t$ means the longest surviving routing cycle
relative to the decadal filtration scale.

The third observable is the \betti{0} component count under the
adaptive filtration threshold. This quantity tracks metric
fragmentation. A rise in \betti{0} indicates that regions of the
network that may remain connected in the graph-theoretic sense are
no longer functionally close under the relevant transport cost
envelope.

\subsection{Structural break tests}
\label{sec:structural_breaks}

To identify statistically significant changes in the topological
time series, the Chow structural break test is used. For a
candidate break point $\tau$, the Chow test compares a single linear
regression fitted over the full time series with two separate
linear regressions fitted before and after $\tau$. The test
statistic is
\begin{equation}
F =
\frac{
\left(
RSS_{\rm full}
-
RSS_{\rm before}
-
RSS_{\rm after}
\right)/k
}{
\left(
RSS_{\rm before}
+
RSS_{\rm after}
\right)/(n-2k)
},
\label{eq:chow}
\end{equation}
where $RSS$ denotes the residual sum of squares,$k$ is the number if regression parameters (in our case $k=2$) and $n=41$.

The candidate break points are specified from the historical
record 
The principal tests are performed at 260\,\ce,
corresponding to the peak of the Crisis of the Third Century, and
at 310\,\ce, which is usually associated with the transition into the late
imperial structural regime. This historically constrained
procedure avoids the inflation of significance that would arise
from unconstrained changepoint search.

\subsection{Bootstrap validation and permutation test}
\label{sec:bootstrap_validation}

The stochastic component of the friction model introduces
run-to-run variability in edge weights and therefore in the
topological observables. To quantify this variability, the full
pipeline is repeated over independent random seeds. For each
decade, the bootstrap mean and confidence interval of $H_t$ are
computed from the ensemble of runs.

The bootstrap  tests
whether the observed topological phases are robust to plausible
cost fluctuations. Secondly, it prevents individual stochastic
realisations from being overinterpreted as historical signals.
The structural break tests are performed on the bootstrap mean
series,while the confidence intervals determine whether the signal-to-noise ratio of each phase transition is sufficient for structural interpretation.

In addition to the bootstrap, a permutation test was performed
to assess whether the Phase~III slope of $H_t$ is distinguishable
from a null distribution obtained by randomly reassigning decadal
labels ($n=100$ permutations). The resulting test statistic
($s_{\rm obs}=-0.070$) is less extreme than the permutation null
mean ($\bar{s}_{\rm null}=-0.094$, $\sigma_{\rm null}=0.030$,
$p=0.77$). This result must be interpreted carefully. The
permutation test and the Chow break test answer different
questions: the Chow test asks whether the \emph{transition into}
Phase~III represents a statistically significant change of regime
(it does, $F=85.4$, $p<0.001$) and the permutation test asks
whether the \emph{absolute magnitude} of the Phase~III slope is
significantly steeper than a randomly permuted slope (it is not,
$p=0.77$). Both results are consistent: there is robust evidence
of a structural break at 310\,\ce{}, but the post-break rate of
entropy decline is moderate. 

\subsection{East--West partition}
\label{sec:east_west_partition}

The East--West comparison is performed by partitioning nodes at
longitude $20^{\circ}$E. Nodes east of this meridian are assigned
to the Eastern Mediterranean sub-network; nodes west of it are
assigned to the western sub-network.

The eastern network is the primary object of analysis. The western
component is used as a control, but with an important limitation:
the western coverage in the available \orbis{} extract is
incomplete. Therefore, the absence of persistent \bettione{}
homology in the western component 
has to be understood as a
dataset-specific limitation that motivates future extension to a full
Roman--Byzantine network.


\section{Results}
\label{sec:results}

\subsection{Three structural phases in the Eastern Mediterranean network}
\label{sec:three_phases}

The \bettione{} persistent entropy time series of the Eastern
Mediterranean network exhibits three distinct phases between
0 and 400\,\ce. These phases are visible in the temporal trajectory
of $H_t$ and are supported by structural break tests at 260 and
310\,\ce. The sequence is not a monotonic decline from the early
Empire to late antiquity. It is instead a three-regime process:
stability, recoverable crisis, and irreversible structural decline.

\paragraph{Phase I: topological stability, 0--200\,\ce.}

During the first two centuries of the series, the Eastern
Mediterranean network remains in a high-entropy regime. The
persistent entropy is stable, with no
statistically meaningful long-term slope. This indicates that the
network preserved a broad distribution of routing cycles across
the early imperial period. Multiple alternative paths connected
the major commercial and administrative regions of Egypt, the
Levant, Anatolia, Greece, and the eastern frontier.

This result is consistent with the historical interpretation of
the early imperial and Antonine periods as an era of high
Mediterranean integration. 
In
topological terms, redundancy was distributed. No single routing
structure dominated the system. This distribution of cycle
persistence is  what one expects from a resilient
transport network capable of absorbing local disruptions.

A transient entropy depression is observed within the late second
century, corresponding to the period of the second-century pandemic and
military pressure along the imperial frontiers. However, the
network recovers within the following decades. The important point
is not that the early imperial network was unaffected by stress,
but that stress did not produce a lasting change in its
topological regime.

\paragraph{Phase II: recoverable stress, 210--280\,\ce.}

The Crisis of the Third Century produces the first major
topological depression in the series. Persistent entropy declines
around the middle of the third century, with a 
significant structural break near 260\,\ce. This decline coincides
with the period of civil war, imperial fragmentation, plague,
frontier instability, and severe disruption to the eastern
provinces.

The topological signature of this phase is not terminal collapse.
By 280\,\ce, after the Aurelian
reunification of the empire and the restoration of major lines of
communication, the entropy returns to the earlier baseline. This
recovery is the decisive observation in the series. It demonstrates
that the Eastern Mediterranean network still possessed enough
redundant structure to rebuild its cycle architecture after severe
stress.


\paragraph{Phase III: irreversible structural decline, 290--400\,\ce.}

The third phase begins after the recovery from the Crisis of the
Third Century. From approximately 290--310\,\ce{} onward, the
persistent entropy enters a sustained decline. Unlike the
third-century depression, this decline does not reverse before
400\,\ce. The Chow test at 310\,\ce{} identifies the strongest
structural break in the series, indicating that the network
entered a new regime.

The historical significance of this result lies in its timing. The
decline occurs during and after major political and administrative
efforts to restore imperial coherence. The Diocletianic reforms
reorganised taxation, administration, and military command.
Constantine reunified imperial authority after the Tetrarchic
conflicts. Yet the topological series shows no return to the
Phase~I regime. Political consolidation did not restore the earlier
geometry of exchange.

Phase~III is therefore interpreted as an irreversible structural
decline in the network's recovery capacity. The system did not
simply become more expensive to operate; it lost the diversity of
cycles that made recovery possible. The Eastern Mediterranean
network remained active, but its redundancy became increasingly
concentrated and fragile.

Table~\ref{tab:phases} summarises the quantitative properties of
the three structural phases.

\begin{table}[htbp]
\caption{Summary statistics of the three structural phases in
\bettione{} persistent entropy (bootstrap mean series). Slope and
$R^2$ are from ordinary least squares regression over each phase
interval. $H_{\min}$ and $H_{\max}$ are the bootstrap-mean
extrema within each phase; IC$_{95}$ width is the mean 95\%
bootstrap confidence-interval width over the phase.}
\label{tab:phases}
\begin{ruledtabular}
\begin{tabular}{lccccccl}
Phase & Period & $n$ & Mean $H$ & $H_{\min}$ & $H_{\max}$
      & Slope (yr$^{-1}$) & Interpretation \\
\hline
I: Stability  & 0--200\,\ce   & 21 & 3.378 & 3.322 & 3.426
& $-2.7\times10^{-6}$ & Stationary \\
II: Crisis    & 210--280\,\ce &  8 & 3.365 & 3.152 & 3.426
& $-3.0\times10^{-4}$ & Recoverable \\
III: Decline  & 290--400\,\ce & 12 & 3.206 & 3.088 & 3.413
& $-2.2\times10^{-3}$ & Irreversible \\
\end{tabular}
\end{ruledtabular}
\end{table}

\begin{figure}[htbp]
\centering
\includegraphics[width=\textwidth]{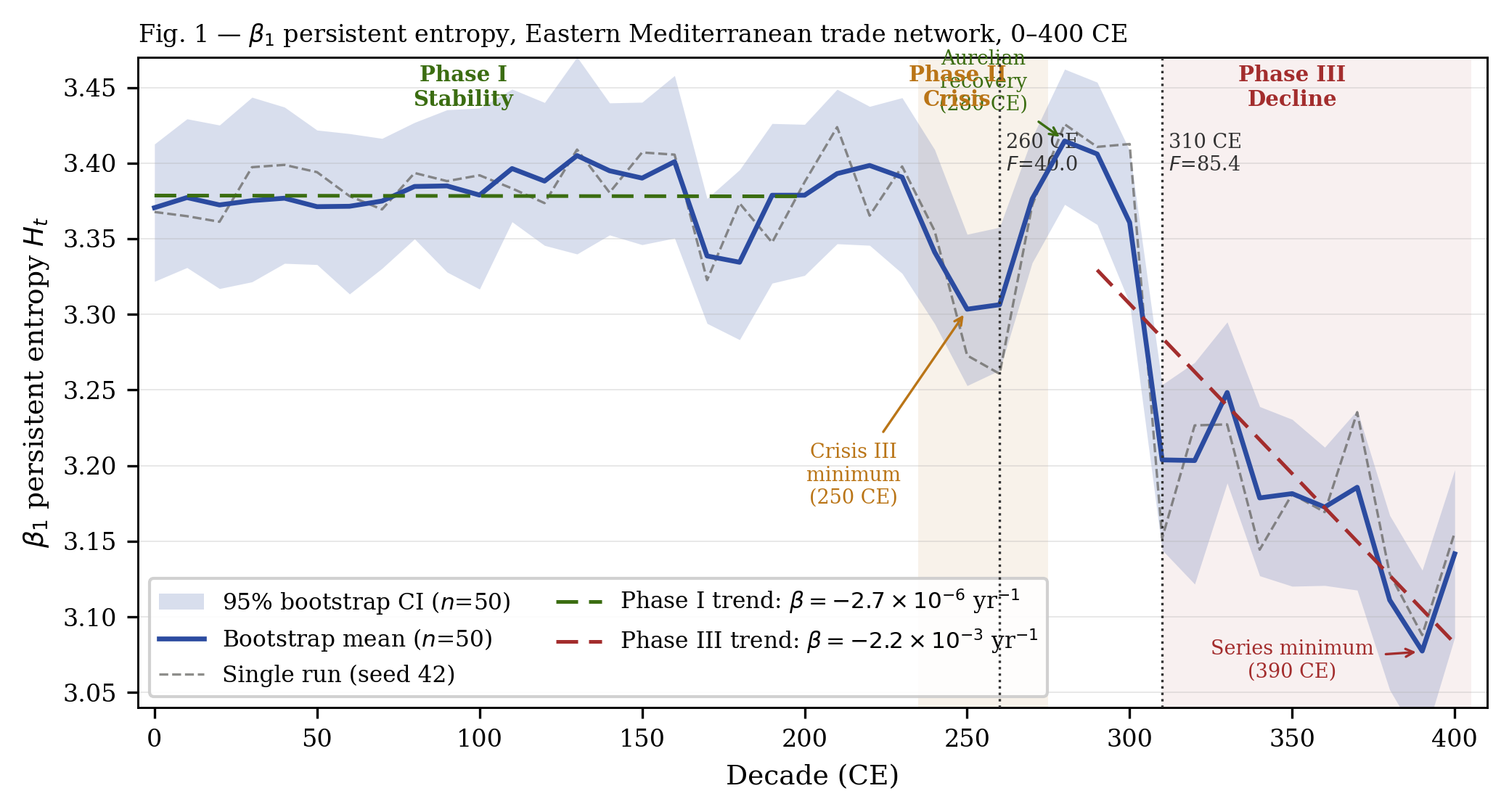}
\caption{\bettione{} persistent entropy of the Eastern
Mediterranean trade network, 0--400\,\ce{} (bootstrap mean, 95\%
confidence interval shaded). Three structural phases are separated
by breaks at 260\,\ce{} and 310\,\ce. The Phase~III decline is
substantially steeper than the Phase~I trend and proceeds without
recovery after the early fourth-century transition.}
\label{fig:main_entropy}
\end{figure}

\subsection{Structural breaks at 260 and 310\,\ce}
\label{sec:break_results}

The Chow structural break tests confirm that the changes observed
in the entropy series are not merely visual fluctuations.
Table~\ref{tab:chow} reports the $F$-statistics and $p$-values
for the seven candidate break points tested. The break at
260\,\ce{} ($F=40.0$, $p<0.001$) corresponds to the peak of
the Crisis of the Third Century and marks the transition into
the deepest part of Phase~II. The break at 310\,\ce{} is
stronger ($F=85.4$, $p<0.001$) and marks the onset of Phase~III.

\begin{table}[htbp]
\caption{Chow structural break test results for the Eastern
Mediterranean \bettione{} persistent entropy series. Tests are
performed at historically motivated candidate break points only.
Significance levels: $*\,p<0.05$; $**\,p<0.01$; $***\,p<0.001$.
The two principal breaks discussed in the main text are shown in
bold.}
\label{tab:chow}
\begin{ruledtabular}
\begin{tabular}{ccc}
Break point & $F$-statistic & Significance \\
\hline
165\,\ce{} (Antonine Plague onset)          &  \phantom{0}3.1 & ---  \\
193\,\ce{} (Year of Five Emperors)          &  \phantom{0}5.8 & *    \\
249\,\ce{} (Plague of Cyprian onset)        &  12.3           & **   \\
\textbf{260\,\ce{} (Crisis III peak)}       & \textbf{40.0}   & ***  \\
\textbf{310\,\ce{} (late imperial onset)}   & \textbf{85.4}   & ***  \\
340\,\ce{} (Gothic pressure onset)          &  \phantom{0}6.2 & *    \\
378\,\ce{} (post-Adrianople)                &  \phantom{0}3.9 & *    \\
\end{tabular}
\end{ruledtabular}
\end{table}

The 260\,\ce{} break is associated with a large
but recoverable disruption. The 310\,\ce{} break is associated
with a smaller immediate political drama but a more important
structural outcome: the disappearance of recovery. The network
survived the third-century crisis as a topological system. It did
not return to its previous regime after the early fourth-century
transition.

This result shifts the focus from collapse as an event to collapse
as a loss of resilience. A system may remain politically
recognisable after its recovery capacity has already been damaged.
In that sense, the fourth-century topological decline is a
precursor to later imperial fragmentation rather than a simple
reflection of it.

\subsection{Western sub-network result}
\label{sec:western_result}

The western sub-network available in the \orbis{} extract does not
produce persistent \bettione{} homology during the study period.
Its persistent entropy remains zero because no one-dimensional
cycles survive the filtration at the relevant thresholds. In purely
topological terms, the available western component lacks measurable
cycle redundancy. The structural asymmetry between the two
sub-networks is substantial, as shown in Table~\ref{tab:ew_compare}.

\begin{table}[htbp]
\caption{Structural comparison of the Eastern and Western
sub-networks, averaged over the full study period 0--400\,\ce.
The eastern values correspond to the single-run (seed 42) series;
eastern extrema are bootstrap-mean values.}
\label{tab:ew_compare}
\begin{ruledtabular}
\begin{tabular}{lccc}
Quantity & Eastern network & Western network \\
\hline
Core nodes (mean)         & 551  & 8    \\
\betti{0} (mean)          & 13.2 & 29.3 \\
\bettione{} entropy $H_t$ & 3.09--3.43 & 0.0 (all $t$) \\
\bettione{} features (mean $m$) & 68.8 & 0    \\
\end{tabular}
\end{ruledtabular}
\end{table}

This result has two possible interpretations. The first is
historical: the western commercial system represented in the
extract may have been structurally weaker, less redundant, and
more locally organised than the Eastern Mediterranean network.
This would be consistent with archaeological evidence for the
relative fragility of western long-distance exchange and the
stronger dependence of western commerce on military and
administrative redistribution.

The second interpretation is methodological: the western coverage
of the available extract is incomplete. The absence of Italy, Gaul,
Britannia, and the Rhine frontier prevents the western graph from
reconstructing the major circuits that would be necessary for a
full test of western resilience. Under this interpretation,
$H_t=0$ is not a property of the historical West but a property of
the dataset.

The present paper adopts the conservative interpretation. The
western result is reported as a limitation and as a motivation for
future work on a full imperial network. The primary conclusion of
the study concerns the Eastern Mediterranean sub-network, whose
coverage is sufficiently dense to support the topological analysis.

\begin{figure}[htbp]
\centering
\includegraphics[width=\textwidth]{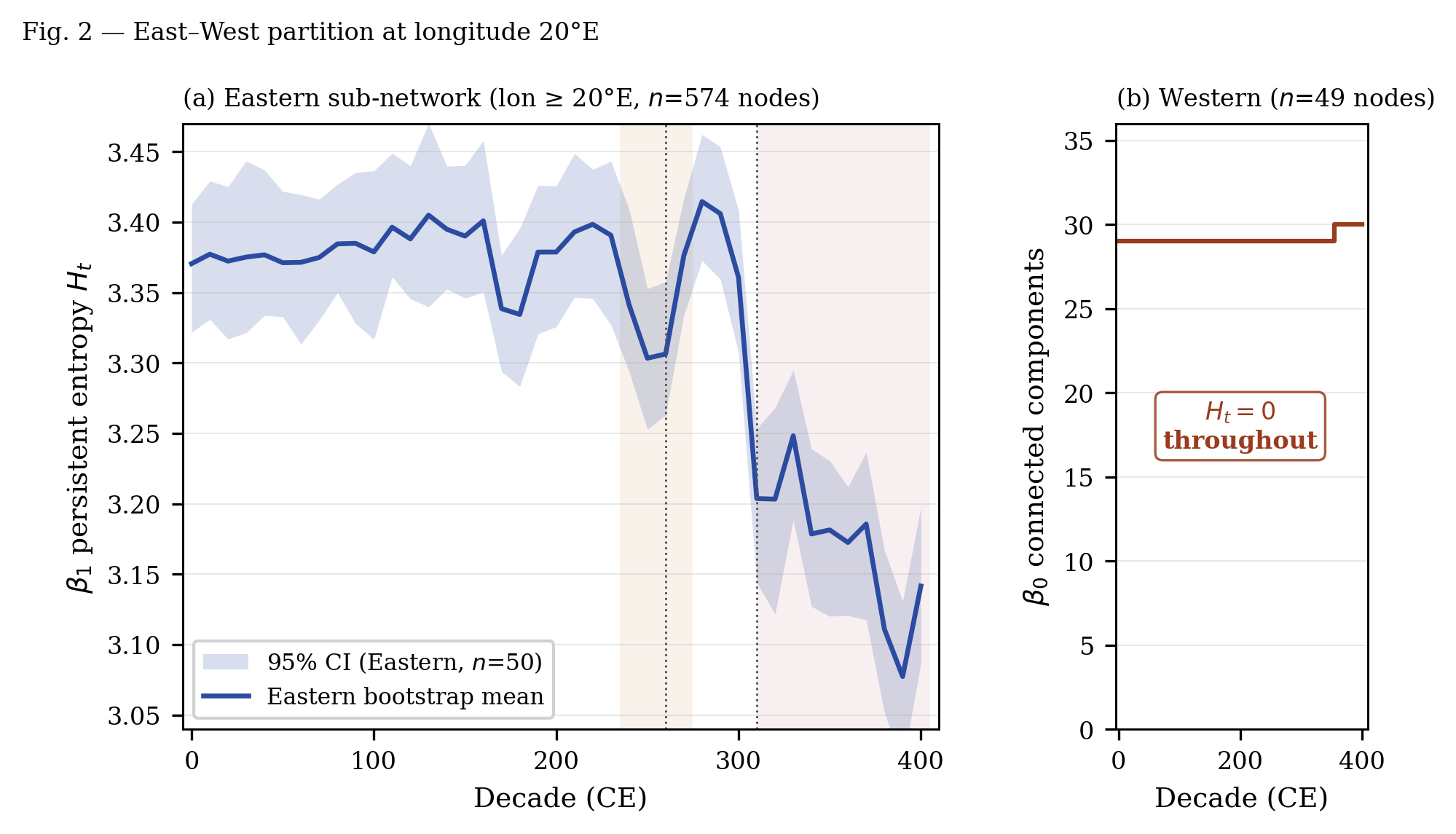}
\caption{East--West partition of the \orbis{} network at longitude
$20^{\circ}$E. The eastern sub-network displays a measurable
three-phase \bettione{} entropy trajectory, while the western
sub-network available in the extract produces $H_t=0$ throughout
the study period.}
\label{fig:east_west}
\end{figure}

\subsection{Military withdrawal and early-warning behaviour}
\label{sec:military_warning}

The military presence index provides an additional test of the
relation between institutional support and network topology. In the
model, military presence affects transport costs by reducing the
effective friction of routes near garrisoned or strategically
controlled regions. The removal or weakening of military presence
therefore increases the cost of movement and may reduce cycle
persistence.

The entropy and normalised lifespan series show that some military
changes precede topological deterioration by one or two decades.
This is consistent with an early-warning interpretation
\cite{Scheffer2009}: changes in security and route maintenance may first affect the maximum
persistence of key cycles before they appear as a network-wide
entropy decline.

The departure of Legio~III Gallica from Syria provides one
illustrative case. The military withdrawal precedes a local minimum
in the normalised \bettione{} lifespan, suggesting that the longest
surviving routing cycles in the Syrian and eastern frontier region
became less resilient after the reduction of military support. The
signal is not sufficient by itself to prove causality, but it is
consistent with the broader mechanism encoded in the friction
model: military infrastructure did not only defend territory; it
stabilised the cost geometry of commerce.

\begin{figure}[htbp]
\centering
\includegraphics[width=\textwidth]{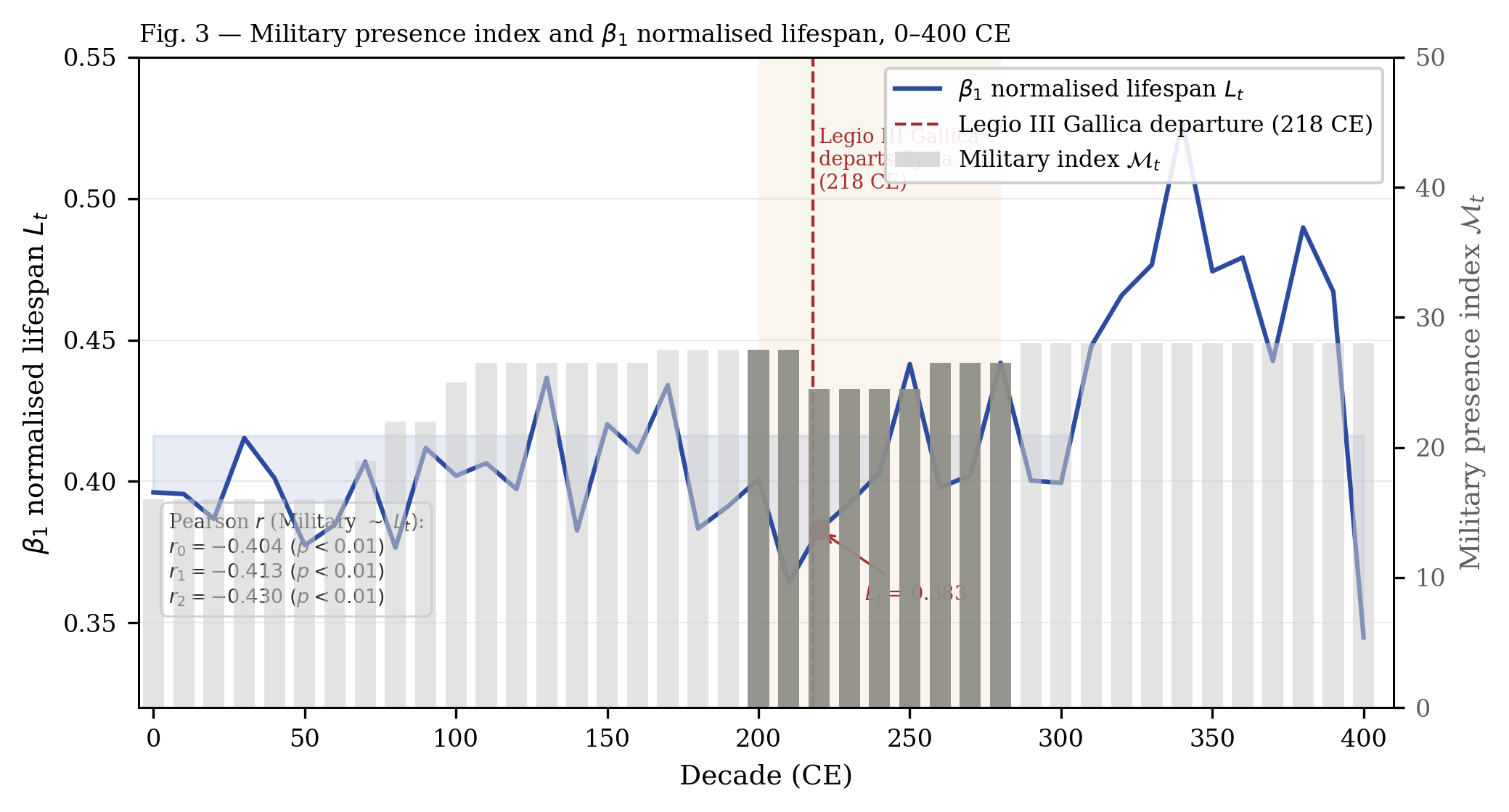}
\caption{Military presence index and \bettione{} normalised
lifespan, 200--300\,\ce. The lagged relation suggests that changes
in military deployment may precede measurable topological
deterioration by one to two decades.}
\label{fig:military}
\end{figure}

\subsection{Late-period fragmentation}
\label{sec:fragmentation}

The late fourth century shows the clearest evidence of combined
cycle loss and metric fragmentation. During Phase~III, \bettione{}
persistent entropy declines while the \betti{0} component count
begins to rise. This simultaneous behaviour is the strongest
topological signature of structural degradation in the series.

In Phase~II, entropy declined but \betti{0} remained stable. The
network lost some cycle diversity but did not fragment under the
adaptive cost threshold. In Phase~III, the loss of cycle redundancy
is accompanied by the emergence of additional components. The
network is therefore not only losing alternative routes; it is
also becoming functionally divided into more weakly connected
regions.

The third-century crisis
is a stress event within a still-connected system. The late fourth-century  is a
coupled process of redundancy loss and fragmentation.

\begin{table}[htbp]
\caption{Active node count and \betti{0} component number in the
eastern core sub-network, 350--400\,\ce. The rise in \betti{0}
coincides with the deepest \bettione{} entropy minima of the
series.}
\label{tab:nodeloss}
\begin{ruledtabular}
\begin{tabular}{cccc}
Decade & Active nodes & \betti{0} & $H_t$ \\
\hline
350 & 550 & 13 & 3.182 \\
360 & 549 & 13 & 3.173 \\
370 & 548 & 13 & 3.186 \\
380 & 545 & 14 & 3.111 \\
390 & 540 & 15 & 3.077 \\
400 & 535 & 15 & 3.142 \\
\end{tabular}
\end{ruledtabular}
\end{table}

\subsection{The Balkanisation Scissors}
\label{sec:balkanisation_scissors}

The joint behaviour of \betti{0} and \bettione{} across the three
structural phases produces what we term the \textit{Balkanisation
Scissors}: a simultaneous increase in connected components
(\betti{0}$\uparrow$) and decrease in cycle redundancy
($H(\beta_1)\downarrow$). This configuration defines a topological fingerprint of structural disintegration in the $(\beta_0, H(\beta_1))$ phase plane.

In Phase~I, the eastern network occupies a compact region of the
$(\beta_0,H(\beta_1))$ plane. \betti{0} remains near its baseline
while $H(\beta_1)$ remains high. Phase~II shifts the trajectory
downward in entropy while leaving \betti{0} stable.  
In Phase~III,  \betti{0} rises while
$H(\beta_1)$ declines,signalling the concurrent erosion of metric connectivity and cycle diversity.

In recoverable stress, entropy may decline but
metric connectivity remains intact. In irreversible decline, the
network loses both alternative cycles and effective connectivity.

The absence of persistent \betti{2} features throughout the study
period is also informative. It indicates that the eastern Roman
trade network is topologically flat at the filtration scales used
here. Its fragmentation occurs in the \((\beta_0,\beta_1)\) plane:
the network loses cycles and breaks into components. 

\begin{figure}[htbp]
\centering
\includegraphics[width=\textwidth]{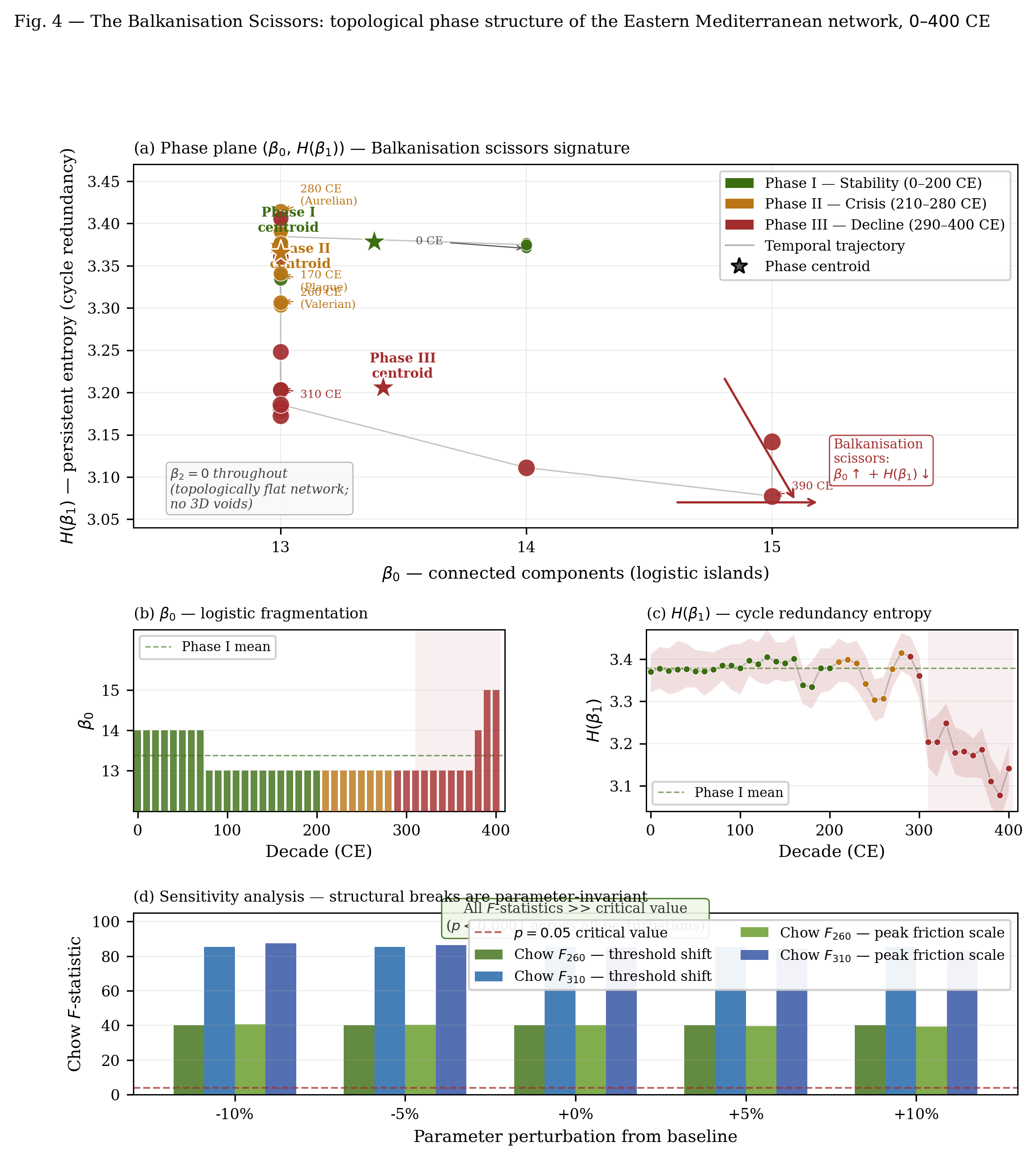}
\caption{The Balkanisation Scissors: topological phase structure
of the Eastern Mediterranean trade network, 0--400\,\ce.
(a) Phase plane \((\beta_0,H(\beta_1))\). Each point is one decade;
colour encodes phase; stars mark phase centroids. The scissors
pattern, defined by simultaneous \(\beta_0\uparrow\) and
\(H(\beta_1)\downarrow\), is activated primarily in Phase~III.
(b) \(\beta_0\) time series. (c) \(H(\beta_1)\) time series with
bootstrap confidence interval. (d) Sensitivity analysis of Chow
statistics under perturbations of filtration and friction
parameters.}
\label{fig:scissors}
\end{figure}


\section{Discussion}
\label{sec:discussion}

\subsection{Recoverable stress and irreversible decline}
\label{sec:recoverable_irreversible}

 The
difference between the third-century depression and the
fourth-century decline is the main result of this analysis.

The third-century crisis produced a measurable reduction in
\bettione{} persistent entropy but  the network succeeded to recover. By the
late third century, the entropy series returned to the early
imperial baseline. This recovery shows that the network 
retained the routing redundancy required to reorganise after
severe disruption. In topological terms, the Crisis of the Third
Century 
did not destroy the
system's capacity to regenerate the damaged cycles.

The post-290\,\ce{} trajectory is different. After the early
fourth-century transition, \bettione{} entropy declined and did not recover. 
The
system remained connected enough to function, but unable of 
absorbing perturbations through alternative routes.

A transition becomes irreversible not at the moment of maximum perturbation, but when the system's recovery capacity falls below the critical threshold required for structural regeneration.The
topological evidence  places the decisive transition not
at the point of greatest immediate disruption, but at the point
where recovery ceases to occur.

\subsection{Topology and political reunification}
\label{sec:topology_politics}

The timing of Phase~III is historically significant 
. The
decline begins during a period in which the imperial state was
attempting to restore administrative and military coherence.
Diocletianic reorganisation and Constantinian reunification
produced real political changes, but these changes do not appear
as a topological recovery in the \bettione{} entropy series.

This result suggests that political reunification and network
restoration were not equivalent processes. The state could reassert
authority over territory without restoring the earlier geometry of
exchange. Administrative reform could stabilise taxation or command
structures while leaving the redundancy of commercial and
logistical routes diminished. In this sense, the network retained
a memory of structural degradation that was not erased by political
consolidation.

This does not  mean  that political events were irrelevant.
Civil wars, reforms, military deployments, and imperial
reunifications all affected the cost structure of the network. After a certain threshold, changes in
political authority no longer restored the lost topological
redundancy. The recovery after Aurelian and the absence of a
comparable recovery after Constantine therefore define two
different regimes of imperial resilience.

\subsection{Relation to existing explanations of Roman decline}
\label{sec:historiography}

The results do not support a single-cause explanation of Roman
decline. Instead,  several
historiographical interpretations can be compared within a common structural
formalism.

Military explanations emphasise frontier pressure, the diversion
of resources to defence, and the declining capacity of the Roman
army to secure territory and infrastructure. The topological
results are consistent with this view insofar as military presence
appears to stabilise transport costs and route persistence. The
possible lag between military withdrawal and topological
deterioration suggests that the army functioned  as an infrastructure of commercial
security. However, military pressure alone does not explain the
entire entropy trajectory.

The transient entropy decline associated with the
Antonine Plague and the deeper third-century disruption are
compatible with demographic-shocks interpretation. Epidemic mortality could
reduce demand, labour availability, transport capacity, etc. Yet the network recovered from earlier biological and
demographic shocks. This means that epidemic stress was important,
but not sufficient to explain the irreversible character of
Phase~III.

Economic transformation provides the closest match to the
topological evidence. The transition from an integrated
tax-and-redistribute system toward more localised exchange should
appear precisely as a decline in long-distance cycle redundancy.
A network with fewer persistent \bettione{} cycles is one in which
fewer alternative long-distance circuits remain viable. The rise in
\betti{0} fragmentation during the late period further supports the
interpretation that some regions became functionally more isolated
under the operative cost scale.

The topological evidence therefore does not replace existing
historical explanations. It reorganises them. Military pressure,
epidemic mortality, fiscal stress, and political instability are
not independent causes acting on an inert system. They are
perturbations acting on a network whose capacity for recovery
changes over time. Persistent homology measures that capacity
directly.

\subsection{The meaning of \texorpdfstring{\(\beta_1\)}{beta1}
entropy in historical networks}
\label{sec:meaning_entropy}

The interpretation of \bettione{} persistent entropy requires
care. 
It means that the
lifespans of one-dimensional topological features were distributed
across multiple cycles. In a transport network, this
corresponds to a system with several alternative routing structures
of comparable persistence.

Similarly, a decline in $H_t$ 
means that commerce could continue through a smaller
number of robust corridors. Indeed, a network may remain
commercially active while becoming topologically fragile. 
If redundancy is carried by fewer
cycles, then the failure or cost inflation of one corridor has
larger systemic consequences.

This is why persistent entropy is useful for studying imperial
stress. 
A scalar economic index may
show that costs increased or that trade volume decreased.
Persistent homology shows whether the geometry of alternative
routes remained diverse enough to sustain the operational integrity of the network.

\subsection{The western sub-network as a limitation}
\label{sec:west_limit}

The western result must be interpreted conservatively. In the
available \orbis{} extract, the western sub-network does not
produce persistent \bettione{} homology. 

The limitation is geographic coverage. A complete western analysis
would require Italy, Gaul, Britannia, the Rhine frontier, North
Africa, and Iberia to be represented with node density comparable
to that of the Eastern Mediterranean. Without that coverage, the
western graph is structurally under-sampled.


\subsection{Methodological contribution}
\label{sec:method_contribution}

The construction
of a pipeline for analysing historical transport networks through
persistent homology is our main methodological result. The pipeline has four components.

\begin{itemize}
    \item it converts a static historical network into a temporal
sequence of weighted graphs. This is necessary because historical
stress often changes the cost of movement before it removes nodes
or edges. A road, port, or maritime corridor may continue to exist
physically while becoming functionally less viable.
\item  the model uses differential friction in place to use uniform
scaling. This is important because transport modes respond
differently to historical perturbations. 
\item the analysis uses an adaptive filtration threshold. The
90th-percentile threshold avoids overemphasising extreme distances
and makes the filtration sensitive to the functional core of each
decadal network where  peripheral nodes and uncertain edges can
dominate the persistence diagram.
\item  structural break testing to topological
time series is applied. This moves the interpretation of persistence diagrams away from visual inspection and toward statistical phase
identification. The observables as  persistent entropy and statistical analysis observables as  bootstrap
confidence intervals, and Chow tests allow us to distinghish  ordinary fluctuation from  recoverable
stress or  irreversible decline.
\end{itemize}

\subsection{Limitations}
\label{sec:limitations}

Several limitations should be made explicit.
\begin{itemize}
    \item First,  the quality and coverage of the
\orbis{} dataset are strongly relevant for this analysis. The Eastern Mediterranean is well represented
relative to the western component, but the network remains a
reconstruction. Node dates, route costs, and transport
classifications contain uncertainty. The bootstrap procedure
addresses stochastic uncertainty in edge weights,the uncertainty on the underlying historical dataset is still there.

\item Second, the edge infrastructure is treated as topologically static
within the study interval. Edges change weight through the friction
model, but routes are not dynamically removed except through node
activation. 
A more complete model would include edge activation and
deactivation based on archaeological or textual evidence.

\item Third, the friction multipliers are historically motivated but not
uniquely determined. Different calibrations could change absolute
entropy values. 
Future work should test broader sensitivity ranges.
\item Fourth, the temporal resolution is decadal. This scale is
appropriate for long-term structural analysis, but it cannot
resolve short-lived responses to individual events. Some events
that occurred within the same decade are necessarily aggregated.
Annual or sub-decadal analysis would require a denser and more
reliable temporal dataset.
\item Fifth, the military presence index is incomplete. The Roman army
was spatially complex, and its relationship to trade security was
not uniform across provinces. The present model captures only a
simplified version of military support. 
\item Sixth, the permutation test ($p=0.77$, Sec.~\ref{sec:bootstrap_validation})
indicates that the Phase~III slope magnitude is not unusual
relative to a randomly permuted null. This constrains the
strength of the irreversibility claim: the evidence supports a
change of structural regime (Chow $F=85.4$) but not an
exceptionally steep rate of post-break decline. 
\end{itemize}

\section{Conclusions}
\label{sec:conclusion}

This study applied persistent homology to the Eastern Mediterranean
network represented in the \orbis{} Geospatial Network Model
of the Roman World between 0 and 400\,\ce. By converting the
static transport network into a sequence of decadal weighted graphs
through a differential friction model, we measured the evolution of
cycle redundancy, maximum cycle persistence, and metric
fragmentation across four centuries of Roman imperial history.

The main result is the identification of three structural phases in
the \bettione{} persistent entropy time series. Phase~I, from
0 to 200\,\ce, is a stable high-redundancy regime. Phase~II, from
210 to 280\,\ce, corresponds to the Crisis of the Third Century as
a recoverable stress event. Phase~III, from 290 to 400\,\ce, is an
irreversible decline in cycle redundancy. An important result is the distinction between
Phase~II and Phase~III where  the network recovered
from the third-century crisis but did not recover after the early
fourth-century transition.

Chow structural break tests identify significant breaks at
260\,\ce{} and 310\,\ce. The first corresponds to the peak of
the third-century crisis. The second marks a stronger transition
($F=85.4$, $p<0.001$) into a regime of sustained topological
decline. A complementary permutation test ($n=100$) shows that
the absolute magnitude of the Phase~III slope is not
significantly steeper than a randomly permuted baseline
($s_{\rm obs}=-0.070$, $p=0.77$); the structural irreversibility
claim therefore rests on the change of regime detected by the
Chow test rather than on the slope magnitude alone.

The analysis also shows that political reunification did not
necessarily imply topological restoration. The entropy decline of
Phase~III continues despite the administrative and political
consolidation associated with Diocletian and Constantine. This
indicates that the geometry of exchange retained structural damage
that was not reversed by imperial reunification.


In conclusion, our work shows that  imperial decline should be studied
not only through events, institutions, or scalar economic
indicators, but also through the topology of the networks that
sustained the imperial system. Persistent homology detects the loss
of route redundancy and the emergence of metric fragmentation
before these processes necessarily appear as complete political
collapse. In this sense, the relevant collapse variable is not
stress alone, but the disappearance of the network's capacity to
recover from stress.


\begin{acknowledgments}
The authors thank the Stanford Humanities + Design Lab for making
the \orbis{} dataset publicly available, and the contributors to
the Pleiades and \textsc{dare} gazetteers for their open-access
temporal data. The \orbis{} source data are available at
\url{https://orbis.stanford.edu}. We acknowledge financial support from SECIHTI and SNII (M\'exico).
\end{acknowledgments}





\end{document}